# MORIOND QCD 2013 EXPERIMENTAL SUMMARY

Dmitri Denisov

*Fermilab, Batavia, IL 60510 USA*

The article presents experimental highlights of Moriond 2013 QCD conference. This was fantastic conference and the first Moriond QCD since the discovery of the Higgs boson. Many new results about its properties have been presented at the conference with "Higgs-like" particle becoming "a Higgs" as it properties match expected for the Higgs boson pretty well. There were many new results presented in all experimental areas including QCD, elecroweak, studies of the top, bottom and charm quarks, searches for physics beyond Standard Model as well as studies of the heavy ion collisions. 56 experimental talks have been presented at the conference and it is impossible to cover each result in the summary, so highlights are limited to what I was able to present in my summary talk[1] presented on March 16 2013. The proceedings of the conference cover in depth all talks presented and I urge you to get familiar with all of them. Theoretical Summary of the conference was given by Michelangelo Mangano, so theory talks are not covered in the article below.

## 1 Introduction

2013 Moriond QCD conference had many important experimental results presented. Many of them contribute to our understanding of the particle physics deeply from the extremely precise measurements of the fundamental constants to hints of new physics effects and phenomena. The results presented in most cases were obtained by large experimental collaborations and I would like to congratulate all of them with providing great advances for the experimental particle physics. All speakers did an excellent job with presentations in some cases presenting results which have been approved just a few hours before the talk. All of the above made 2013 Moriond QCD conference very exciting with talks eagerly awaited and many interesting and productive discussions during the conference.

The article is divided by the main topics which coincide with general themes of the conference established by the conference organizers: QCD, heavy ion physics, new physics searches, heavy flavor studies, studies of the top quark and at the end I'll summarize most recent results on a Higgs boson. As with all conferences by the time the proceedings are published many experimental results are superseded and/or improved. My summary will follow closely actual results and plots presented at the conference, so for the updates readers should go to the experimental collaborations Web sites, arXiv, and the refereed journals.

**2 QCD**

Parton scattering with production of jets is the most copious process at hadron colliders such as LHC and Tevatron. Fig. 1[2] presents general diagram of the scattering process highlighting parameters which could be extracted from the studies of jets. These studies include precision parton distribution functions determination critical for calculations of all cross sections at hadron colliders, measurement of the strong coupling constant as well as measurements of vast array of QCD processes to verify and improve QCD theoretical calculations and search for effects beyond those predicted by the Standard Model.

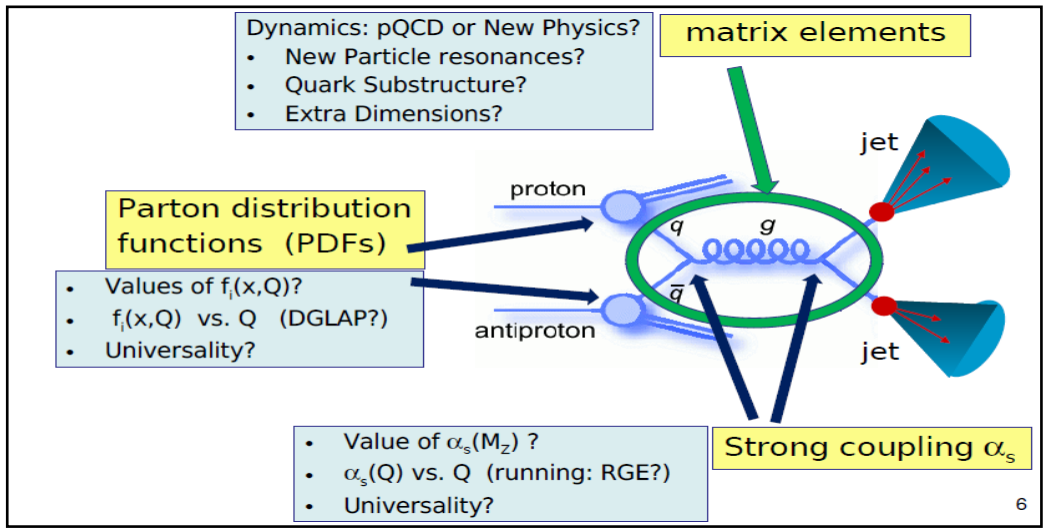

Figure 1. Diagram of a process with the production of a pair of jets and QCD parameters which could be extracted from studies of such process.

Tevatron experiments presented 100's of new measurements from di-jet production to studies of vector bosons production in association with jets. Over many orders of magnitude and up to energies of ~ 1 TeV QCD describes all measured distributions well. Jets studies at the LHC are now probing considerably larger phase space region in both $Q^2$ and values of x as demonstrated on Fig. 2[3]. Presented at the conference di-jet cross sections by CMS are shown on Fig. 3[3]. The span of the cross section measurements covers 10 orders of magnitude and is in excellent agreement with modern QCD predictions in the wide range of jets rapidity and transverse momentum.

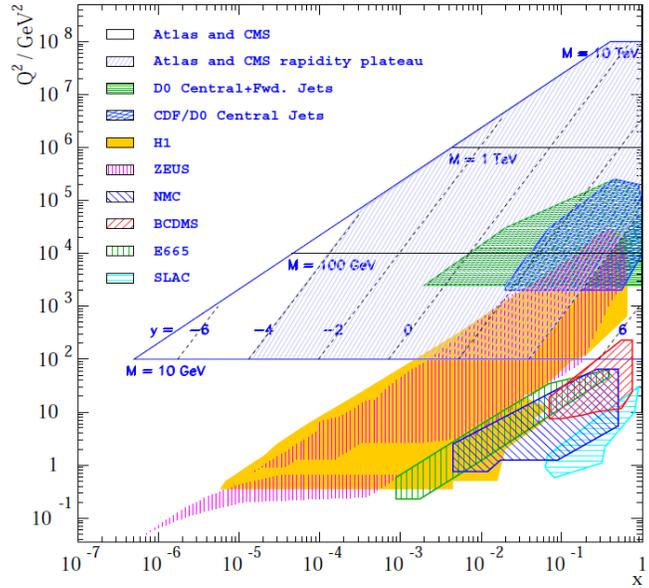

Figure 2. Range in $Q^2$ and x accessible for studies at different accelerators demonstrates unique capabilities of the LHC to study high $Q^2$ and low x processes.

These measurements is a clear indication that in the currently available for experiments kinematical range strong interactions between partons are well described by QCD. Higher energies, including 14 TeV LHC data coming in a few years, will be critical to find if any deviations from the QCD predictions exist at higher energies and accordingly smaller probed distances.

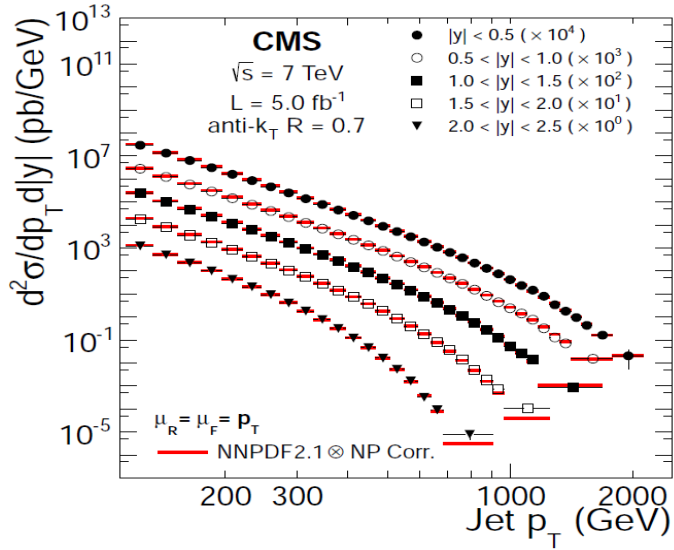

Figure 3. Inclusive di-jet cross section measurements at $\sqrt{s}$ 7 TeV by the CMS experiment at the LHC.

Interesting result presented at the conference was extraction of the strong coupling constant from the ratio of 3-jets to 2-jets cross sections[2]. In such ratio many uncertainties, for example coming from the choice of PDFs, are greatly reduced in comparison with inclusive jet's studies. The results of this measurement provide value of $\alpha_s = 0.1191(+0.0048/-0.0071)$ where main uncertainty is due to factorization and renormalization scales. Same measurement provides first test of strong coupling constant running above ~0.2 TeV: new result from D0 indicates that $\alpha_s$ runs as expected all the way to 0.4 TeV as presented on Fig. 4a[2]. CMS, using higher energy LHC beams and similar method based on ratio of 3-jets to 2-jets cross sections improved Tevatron result as presented on Fig. 4b[3] demonstrating running of the strong coupling constant all the way to 1.0 TeV or ~$2 \cdot 10^{-16}$ cm. This result has fundamental importance as it provides test of QCD up to the smallest distances studied as of today.

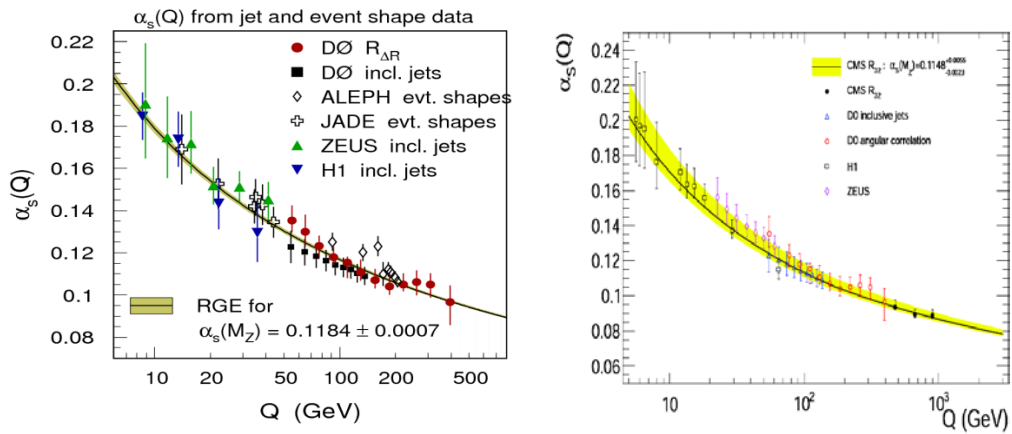

Figure 4. Results on the running of the strong coupling constant at the Tevatron (left, 4a) and at the LHC (right, 4b).

Expected increase in the LHC energy and relatively minor effect on high energy jets from multiple pp interactions will provide test of the running of the strong coupling constant at much smaller distances with LHC 14 TeV data.

Another area where many new results have been presented is studies of direct photons. In addition to testing QCD, probing gluon content of the proton, and understanding background to Higgs studies in two photons decay mode, these results provide unique information about fragmentation as well as test collinear and $k_T$ factorization approaches and soft gluon logarithmic resummation techniques. Fig. 5[4] presents ATLAS results covering direct photon production between 0.1 and 1.0 TeV. Overall there is a good agreement with predictions. Expectation is that these direct photon results will improve gluon PDFs substantially decreasing theoretical error on the Higgs boson production and decay to a pair of photons by as much as 20%. Direct di-photon effective mass spectra are also well described by the theory, except for the masses below ~50 GeV where fragmentation process becomes important.

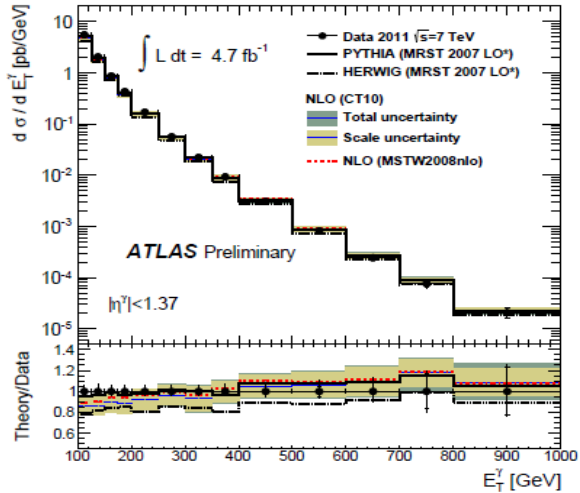

Figure 5. Direct photon production with the ATLAS experiment.

Among process under active experimental scrutiny is W/Z+jets production at the Tevatron and LHC. This process is critical not only for tests of QCD predictions, but as a background to top quark and Higgs boson studies (similar final states) as well as for improvements in PDFs. Large number of kinematic distributions have been obtained providing wealth of information for improvements in theoretical description of these processes. On Fig. 6a ATLAS experiment result[5] for studies of Z boson associated production with up to 7+ jets is presented.

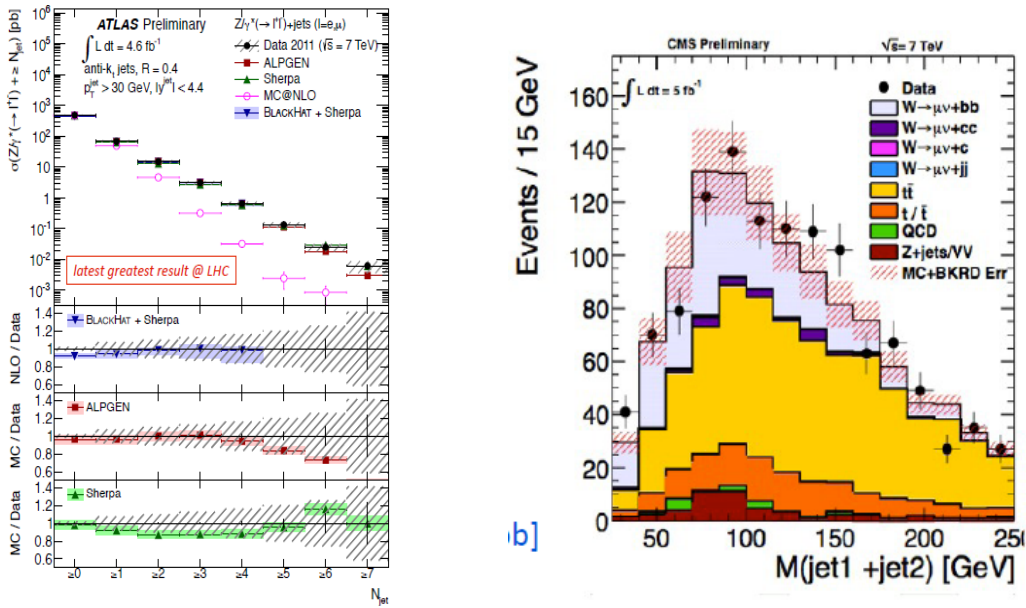

Figure 6. ATLAS studies of Z+jets production (left, 6a) and CMS studies of W+2 b-jets (right, 6b).

Results obtained demonstrate overall agreement with QCD calculations while in some kinematical regions modeling improvements are needed. CMS presented important measurement of W boson production in association with a pair of b-quark jets[5]. Both absolute value of the cross section and shape of the kinematic distributions are in a good agreement with predictions which is important for Higgs boson searches in the decay mode to a pair of b-quarks where associated production of W boson with heavy quark jets is among main backgrounds.

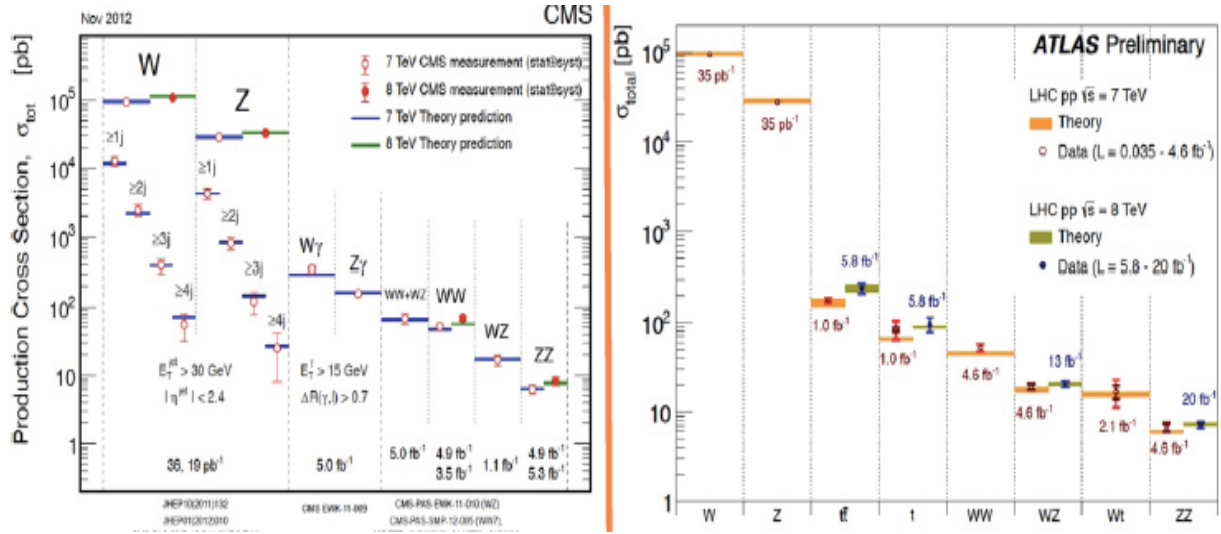

Figure 7. Cross sections of di-boson production as measured by CMS (left) and ATLAS (right).

Di-boson production (W/Zγ, WW, WZ and ZZ) is rare while clean process for tests of electroweak theory and perturbative QCD at TeV scale. These processes are also backgrounds to many channels of searches for physics beyond Standard Model as well as studies of a Higgs boson. Figure 7[6] presents studies of di-boson processes by ATLAS and CMS at the LHC. All cross section measurements as well as kinematic distributions (anomalous triple gauge boson couplings could greatly enhance high energy tails of the distributions) are in good agreement with theory predictions indicating that higher energy data are needed for detection of potential signs of new physics.

## 3 Heavy Ions

Studies of heavy ion collisions at the LHC and RHIC provide critical information to understand behavior of the quark-gluon plasma as well as properties of strongly interacting multiple partons. There is wealth of experimental information obtained recently. LHC provides world highest energy collisions between protons and Pb ions as well as PbPb collisions with up to √s of 2.76 TeV per nucleon. RHIC collides different types of ions with energies in the √s range between 7 GeV and 200 GeV per nucleon. With vast array of data not only properties of the quark-gluon plasma are studied, but the whole evolution of quark-gluon plasma is now understood much better from the initial state conditions (including fluctuations), "perfect" fluid approach, and opaque properties of the plasma where hadronic processes are strongly quenched.

LHC heavy ion collisions provide access to new phase space due to substantially higher energy[7]. Energy density at the LHC reaches ~10 GeV/fm$^3$, volume ~4800 fm$^3$, lifetime 10 fm/$c$ and temperature ~300 MeV. ALICE, ATLAS and CMS presented interesting new results in heavy ion collisions. Below some highlights of these studies are presented. By comparing production of different particles in pp and PbPb collisions important information about quenching of particles produced in quark-gluon plasma is obtained. On Fig. 8[8] results of CMS experiment on $R_{AA}$, or nuclear modification factor, vs type of the secondary particle and its transverse momentum, $p_t$, are presented. $R_{AA}$ compares production of secondary particles in heavy ion collisions with similar final state in pp collisions.

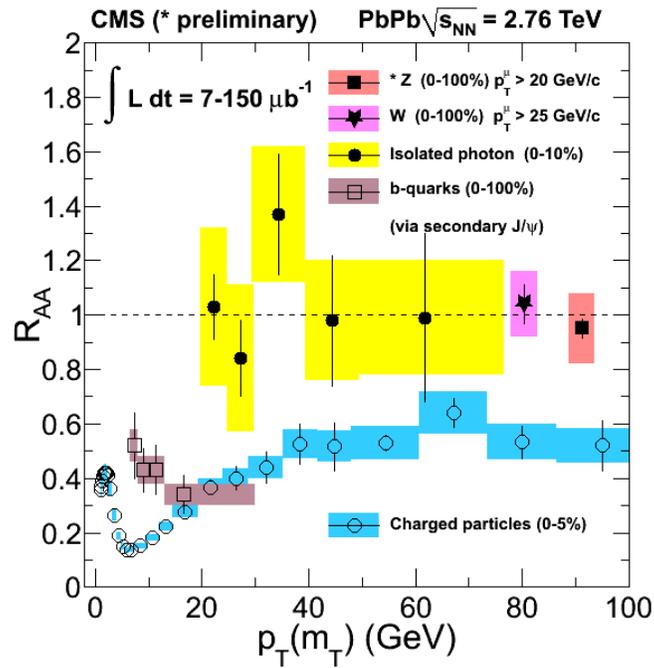

Figure 8. Nuclear modification factor measured for different types of particles by CMS collaboration at $\sqrt{s_{NN}}$ 2.76 TeV.

As clear from Fig. 8 particles which are not affected by strong interaction have $R_{AA}$ close to 1.0 as they propagate through excited nuclear matter of quark-gluon plasma mainly without interactions. But for hadrons $R_{AA}$ is well below 1.0 due to their interaction in the nuclear matter after production. These data are useful for the development of models of quark-gluon plasma.

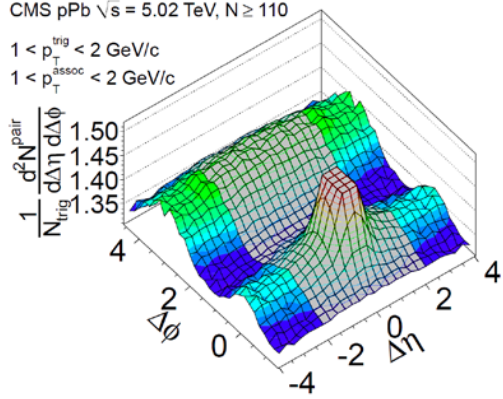

Figure 9. Ridge at Δφ ~0 observed by the CMS collaboration in pPb collisions at √s$_{NN}$ of 5.02 TeV.

Angular correlations between particles produced in heavy ion collisions are among key properties to understand quark-gluon plasma. Near side "ridge" have been observed by CMS[9] in high multiplicity collisions in pPb collisions at √s$_{NN}$ of 5.02 TeV as illustrated in Fig. 9. This effect is similar to the effect observed in pp high multiplicity events and in heavy ion collisions. CMS presented many interesting properties of this effect including dependence on charged particles multiplicity and momentum selection.

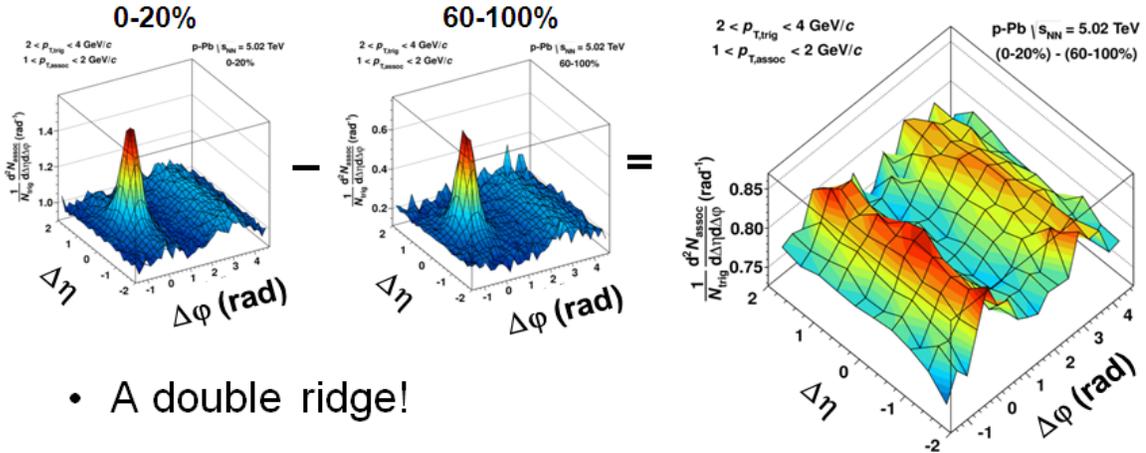

Figure 10. ALICE observation of the "double ridge" structure.

New result[10] was presented by the ALICE experiment at the LHC on angular correlations in pPb collisions and is presented on Fig. 10. All pPb collisions were separated into classes, depending on the charged particles multiplicity with events in 0-20% range been highest multiplicity and in 60-100% range lowest multiplicity. No Δφ ~ 0 ridge is observed in low multiplicity events (similar to pp collisions). But if low multiplicity events distribution is subtracted from high multiplicity "double ridge" structure appears as presented on Fig. 10. Many double ridge properties are studied, while explanation of its origin is yet to come.

Many other exciting results from heavy ion experiments have been presented adding invaluable data to understand dense and hot quark-gluon plasma state of the nuclear matter.

**4 Searches for New Physics**

There are good experimental and theoretical reasons to believe that Standard Model is not complete description of the Nature, so searches for physics effects and new particles beyond those known today is among key priorities for high energy physics. Most natural place to look for new physics is at the energy frontier accelerator where particles with highest masses could be produced. Large number of new physics searches has been presented by the ATLAS and CMS experiments many using full 2012 LHC data set collected at the center of mass energy of 8 TeV.

One of the most attractive new physics models is Supersymmetry (SUSY) which predicts super-symmetric particle for each particle of the Standard Model. While some properties of these particles are predicted, mass scale is not well defined, so searches in the largest accessible phase space are required. ATLAS and CMS experiments both presented many new results at the conference on SUSY searches. Searches are usually divided into strongly produced 1$^{st}$ and 2$^{nd}$ generation scalar quarks and gluinos, 3$^{rd}$ generation scalar quarks (key ingredient in natural SUSY) as well as electroweak production of gauginos and sleptons.

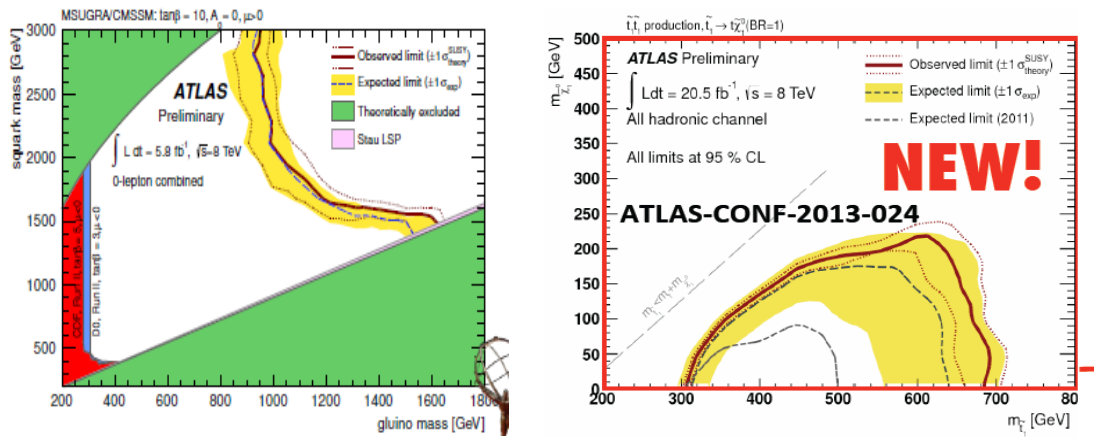

Figure 11. Squark and gluino mass exclusion plot from ATLAS experiment (left plot, 11a). Newest results on stop quarks mass exclusion (right plot, 11b).

Fig. 11a presents ATLAS result[11] in searches for inclusive squarks and gluinos produced via strong interaction. No excess events are observed in the performed searches and very stringent limits on the masses of SUSY particles have been set: up to ~1.6 TeV for gluino and ~3 TeV for squarks. On the Fig. 11b[12] experimental limits on the masses of stop quark by the ATLAS experiment are presented with exclusion now reaching masses in the 600-700 GeV range. Strict limits on SUSY were presented by the CMS collaboration as well[11,12].

An interesting discussion during the conference was devoted to the question "Is SUSY dead?" as searches over ~30 years with ever increasing sensitivity and mass range produced no indication of it existence. In addition to the direct searches there are many indirect searches of SUSY which for now produce "no observation" and in many cases provide strict limits including at high masses. On Fig. 12[13] constrains from direct searches and indirect searches are combined providing (blue points) remaining allowed areas of phase space for CMSSM SUSY particles. It is clearly becoming rather restricted! Conclusions of the discussion about SUSY viability could be summarized as follows:

1. Phase space for SUSY is strongly constrained by direct and indirect searches.
2. Due to ~125 GeV Higgs mass, SUSY particles masses are "multi-TeV" and not yet probed directly.
3. Input from searches not only at accelerators, but, for example, cold dark matter searches, are important to take into account.
4. It is important to continue searches of both direct and indirect manifestations of SUSY particles, especially with LHC energy of 14 TeV.

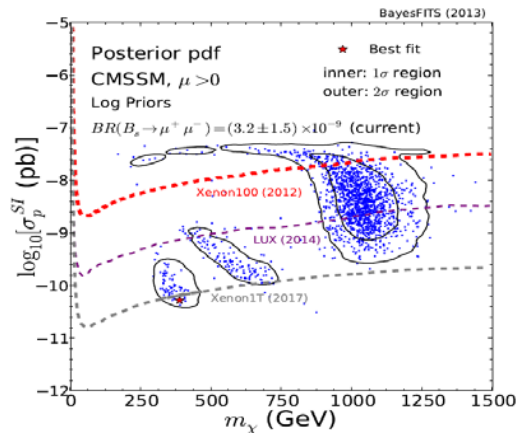

Figure 12. Combination of direct and indirect SUSY (CMSSM version) searches. With expected by 2017 cold dark matter Xenon experiment sensitivity practically all remaining allowed phase space of CMSSM could be excluded.

In addition to "model specific" searches many generic "bump hunting" searches are progressing actively at the LHC. Among those which brought fundamental discoveries in the past (like discovery of the b-quark) are searches for peaks in effective mass spectra of a pair of objects such as leptons, photons or jets. On Fig. 13[14] results of resonance searches in di-electron mass spectra are presented. Excellent energy resolution of LHC detectors and well understood backgrounds provide an opportunity to set limits on such particles as Z' all the way to ~ 3 TeV as no excess events in comparison with Standard Model predictions are observed.

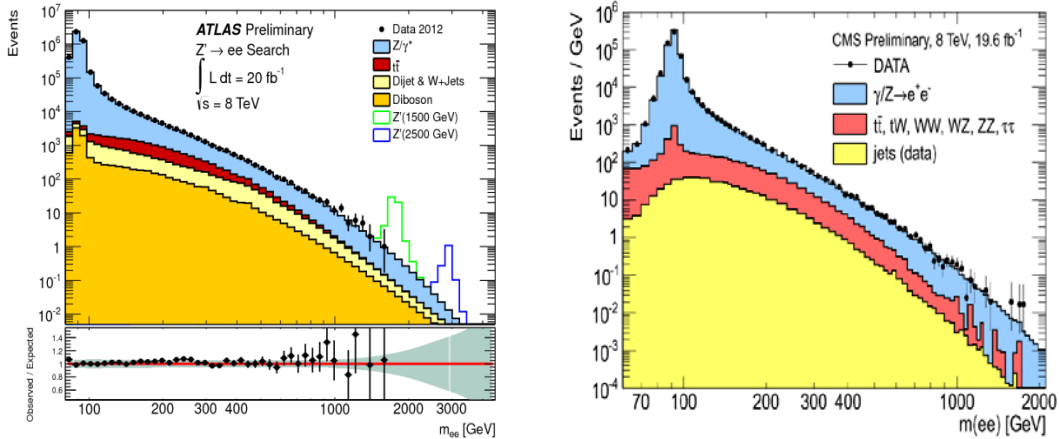

Figure 13. Search for heavy resonances in di-electron spectra by ATLAS and CMS experiments.

The observed di-lepton mass spectra (or similar quantities when lepton and missing transverse energy are studied) are used to set stringent limits on many different models including extra dimensions, excited leptons, heavy W' bosons and others.

Searches for resonances decaying into a pair of jets provide access to the highest masses studied. With available statistics masses up to ~1/2 of the LHC center of mass energy are investigated. On Fig. 14[15] ATLAS and CMS results on searches for di-jets resonances are presented which demonstrate very good agreement with the Standard Model predictions up to the highest masses probed.

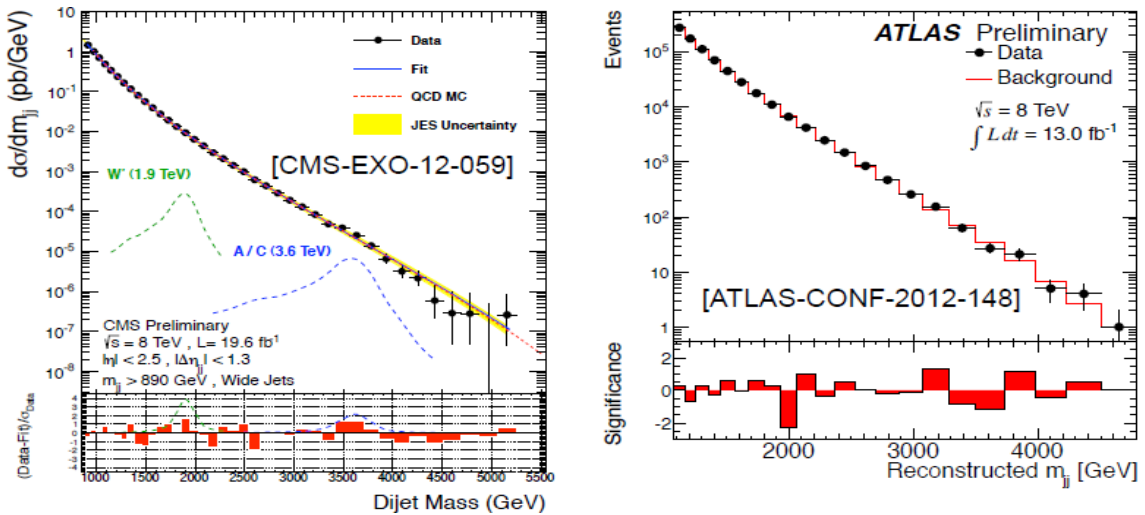

Figure 14. Search for the di-jet resonances by the CMS and ATLAS experiments.

Limits on the mass of the objects decaying to a pair of jets, such as excited quarks, are set at ~ 4 TeV. These searches demonstrate in turn excellent predictions of QCD up to the highest masses studied.

Summary[16] of the LHC experiments searches for new physics are presented on Fig. 15 (ATLAS) and 16 (CMS). With wide array of different searches and masses of objects probed reaching ~10 TeV no hints of beyond Standard Model physics have been seen with 8 TeV LHC data. As major factor in improving reach for new physics (using established methods) is increase in collision energy we will most probably have to wait for 14 TeV LHC data in about 2015 to substantially improve direct searches for new physics at the energy frontier.

There were interesting results on low energy searches for new physics presented at the conference[17,18,19]. They include such important studies as searches for low mass dark matter candidates and rare decays which are enhanced by physics beyond Standard Model. None of these searches indicate excess of events predicted beyond known backgrounds and produce stringent limitations on the theoretical models of new physics.

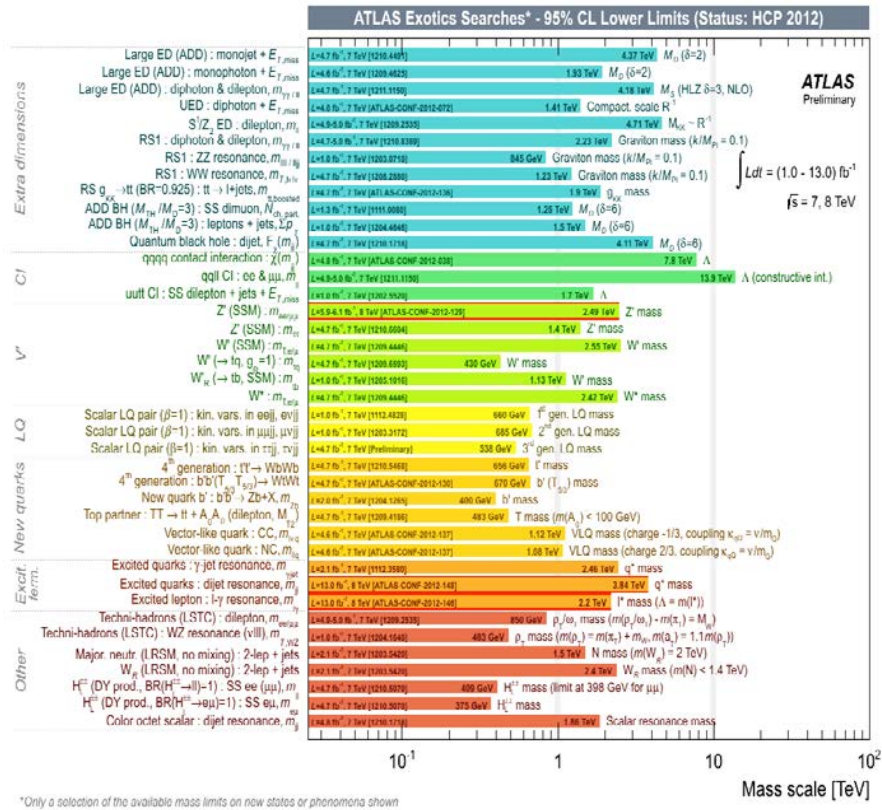

Figure 15. Summary of ATLAS searches for physics beyond Standard Model as of Hadron Collider Physics 2012 conference.

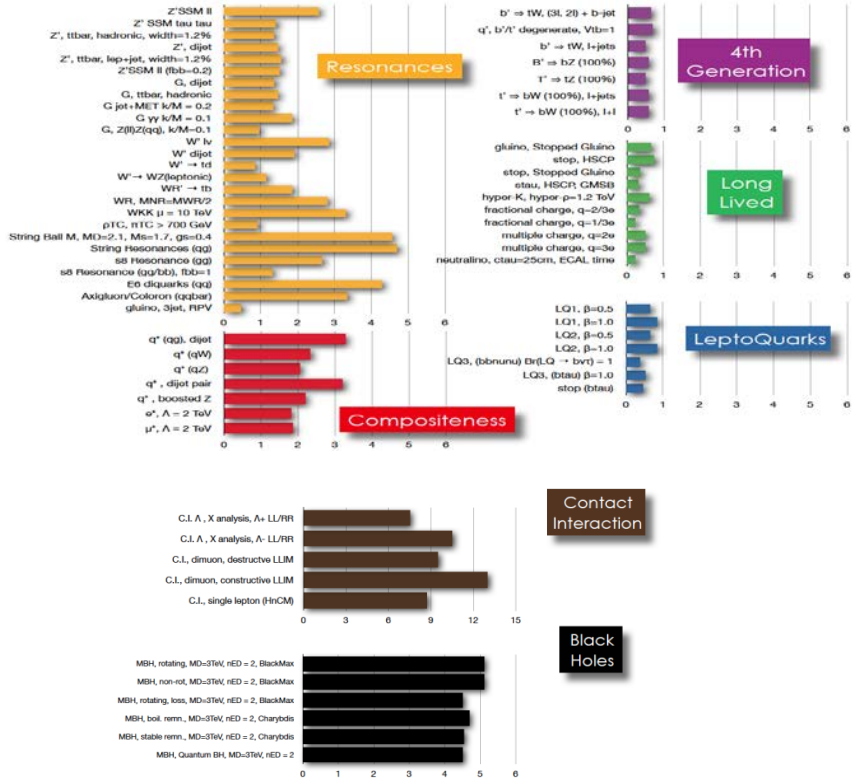

Figure 16. Summary of CMS searches for physics beyond Standard Model as of International Conference on High Energy Physics 2012 conference.

**5 Heavy Flavor Physics**

Studies of particles with heavy c or b-quarks attract a lot of experimental and theoretical attention. Special experiments, like LHCb at the LHC or Belle upgrade at KEK, are designed specifically to study such particles. In depth investigation of the heavy quarks with relatively long life time is important for both our understanding of the Standard Model and for searches for hints of new physics due to quantum effects in flavor loops. Long life time of the c and b-quarks provides an experimental opportunity to identify particles with heavy quarks via secondary decay vertices in high precision silicon detectors. Triggering on events with production and decays of heavy quarks at hadron colliders is usually done via displaced vertex triggers or via soft leptons, mainly muons, which accompany production and/or decay of the particles with c and b-quarks.

Quantum effects in the flavor loops in the presence of new physics, such as SUSY, will affect rates of rare decays. One of the most searched processes is decay of $B_s$ meson to a pair of muons. Any deviation from the Standard Model prediction of $\sim 3 \cdot 10^{-9}$ for the branching fraction will indicate presence of new particles in the quantum loops. Search for this rare decay was progressing for many years with limits decreasing by many orders of magnitude thanks to analyses at both the Tevatron and recently the LHC. LHCb reported first evidence for this rare decay as indicated on the Fig. 17a[20] where mass bump in the effective mass spectrum of pairs of oppositely charged muons is observed.

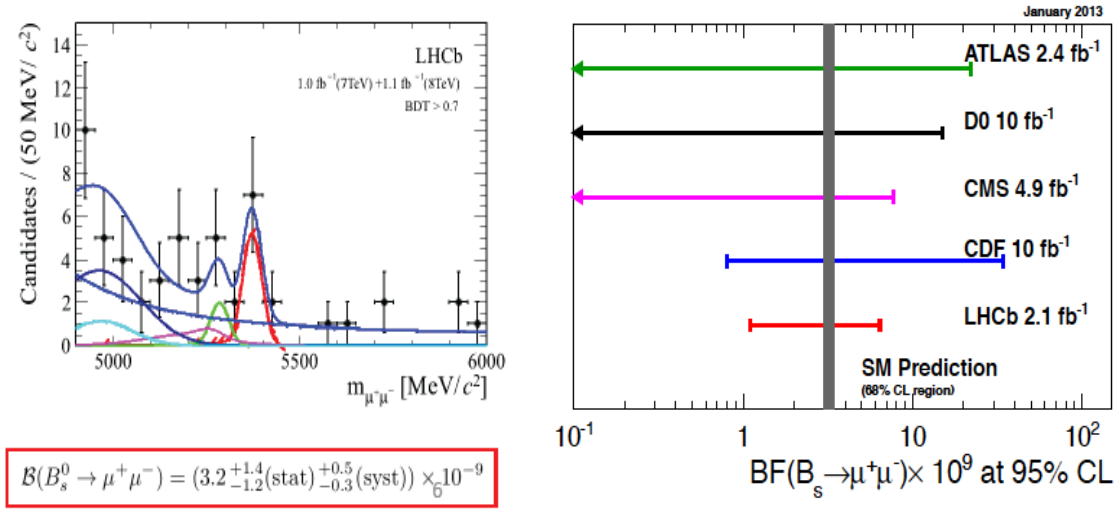

Figure 17. LHCb plot demonstrating evidence for $B_s$ to μμ decay (left, 17a). Compilation of results from different experiments on searches for $B_s$ to μμ decay (right, 17b).

The value of branching fraction observed of 3.2 (+1.4/-1.2) ·$10^{-9}$ (statistical errors which dominate quoted here only) is in a good agreement with the Standard Model predictions further restricting models of the physics beyond Standard Model. Figure 17b[21] presents results of the $B_s$ to μμ decay search by various experiments: all of them are in agreement between each other and with LHCb evidence for this rare process. With more data analyzed by the LHC experiments 5 sigma discovery of this rare decay is expected with full 2012 LHC data set.

Among puzzles in heavy flavor physics is origin of X(3872) particle. All collider experiments see this particle in J/ψππ decay mode, while understanding of its nature is still lacking. In this situation more experimental data are critical and such results are coming from different experiments. CMS experiment presented interesting result studying ratio of X(3872) production (times branching fraction) to ψ(2S) vs transverse momentum of the particles[22]. No significant dependence in the 10-50 GeV range has been observed. LHCb experiment[23] presented studies of spin and parity of X(3872) using angular distributions of the decay products and with high significance determined state as $1^{++}$ which rules out, for example, conventional charmonium state $\eta_{c2}(2^{-+})$. Remaining interpretations of X(3872) involve exotics (non-qq) states.

Belle collaboration reported discovery of new resonances, called $Z_b$, in the decays of Y(5S) to lower energy Y states and a pion[24]. Representative plot of effective mass spectra obtained is presented on Fig. 18a. Belle also provided in depth studies of decay fractions of the new resonances indicating that B(*)B* is the dominant mode of $Z_b$ decays (Fig. 18b).

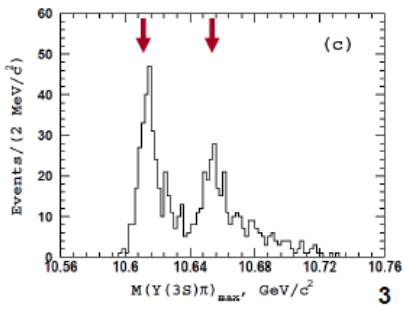

Figure 18. Observation of $Z_b(10610)$ and $Z_b(10650)$ resonances (left, 18a) by the Belle experiment. Decay fractions for the $Z_b$ resonances (right, 18b).

Fig. 18b indicates that this new state is consistent with B*B* molecule providing further information for the development of heavy quark spectroscopy models.

On studies of mesons with b-quarks LHCb collaboration presented extremely precise results on measurements of $B_s$ mixing which is second order weak process sensitive to contributions to beyond Standard Model physics. Using ~34 thousand signal candidate events (1 fb$^{-1}$ of luminosity) mixing frequency was measured with unprecedented precision $\Delta m_s$ = 17.768+-0.023(stat)+-0.006(stat). This measurement is in agreement with theoretical predictions and has accuracy much better than theoretical calculations. Further improvements in the theoretical predictions are welcome while Standard Model stays strong in the heavy flavor sector for now.

While CP violation was observed in the decays of K and B-mesons, no such effect has been observed in the decays of charm mesons and it is expected to be very small in the Standard Model. Recently measurements of CP violation in decays of charm mesons attracted a lot of attention as combined ~4.6σ significance was reported from three experiments: LHCb, CDF and Belle. What is remarkable is the scale of CP violation, about 1%, which is well above Standard Model expectations. Experimental status before Moriond QCD conference is presented on Fig. 19a[25]. As a measure of CP violation the relative difference between number of $D^0$ mesons and anti-$D^0$ mesons decaying to pions or kaons is used. In the directly produced charm decays pion charge is used as a flavor tag, while in the charm from semileptonic B decays muon charge tags flavor. LHCb results[25], presented on a data set of 1 fb$^{-1}$, demonstrate that asymmetry in semileptonic decays is actually positive (while within errors in agreement with zero expected), while in prompt production the updated result moved closer to the "no-CP" region (see Fig. 19b).

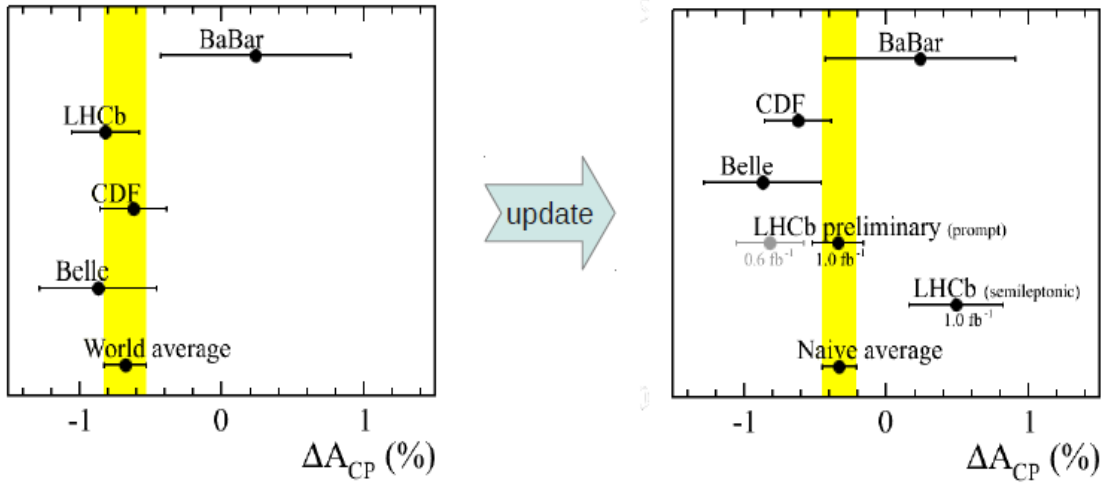

Figure 19. Results on CP violation in charm decays before Moriond QCD conference (left, 19a) and with new results presented by the LHCb experiment (right, 19b).

Results of all experiments are in a reasonable agreement between each other and when combined are within ~2σ compatible with no CP violation in charm decays. While excitement is diminished in this case only analysis of more data will clarify if the original deviation from the Standard Model was a fluctuation or it is a hint of physics beyond Standard Model.

But puzzles are still remaining in the heavy flavor sector. LHCb experiment reported[26] observation of isospin asymmetry in decays of neutral and positive B-mesons well below zero with significance of ~4.4σ, see Fig. 20. While no such asymmetry observed in $K^{*0}/K^{*+}$ decays, in $K^0_s/K^+$ decays asymmetry up to ~50% is observed. There are no natural theoretical explanations for such effect and as this result was obtained on ~30% of the data available most prudent is to wait for the update of this measurement with full 2012 data set.

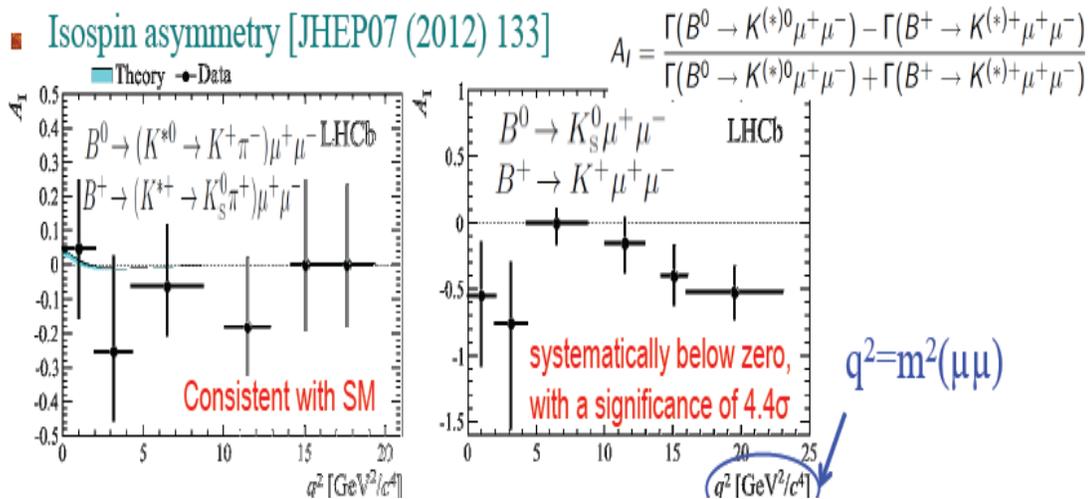

Figure 20. Isospin asymmetry as measured by the LHCb experiment using 1 fb$^{-1}$ of data.

## 6 The Top Quark

There was tremendous progress in our understanding of the heaviest known elementary particle, the top quark, since it discovery which was first reported at Moriond QCD conference in 1995. Precision measurements of the top quark properties are important: to obtain fundamental constants of the Standard Model and as a way to search for physics beyond Standard Model which naturally could demonstrate itself in the properties of the heaviest known particle.

Mass of the top quark is among fundamental parameters and experiments from both LHC and Tevatron presented new measurements in different top quark decay channels. Fig. 21[27] is a summary of the Tevatron top mass measurements updated in March 2013 for Moriond QCD conference.

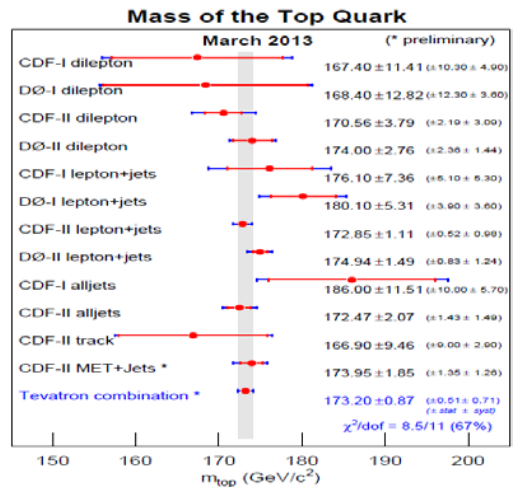

Figure 21. Top quark mass measurements at the Tevatron.

The combined value of the top quark mass is 173.20+-0.87 GeV. The uncertainty reached is 0.5% or twice better in comparison with Tevatron Run II projections. While LHC experiments did not update top mass combination for 2013 winter conferences, summer 2012 result of 173.3+-0.5(stat)+-1.3(syst) GeV is in good agreement with Tevatron measurement and getting similar precision limited now by the systematic uncertainties. It is important to mention that different channels and different methods of measurement of the top quark mass all give consistent results. Still theoretically consistent description of "what mass" is experimentally measured is required for mass measurements with accuracy below ~1 GeV.

CMS experiment at the LHC presented most accurate result on the difference between masses of top quark and top anti-quark[28]. It is -277+-196(stat)+-122(syst) MeV. Within errors the difference is consistent with zero as prescribed by CPT invariance.

Among new measurements in the top quark sector are wide array of top quark pair production cross section measurements including total cross sections and various differential distributions including with the center of mass energy of 8 TeV. On Fig. 22[29] recent 8 TeV results of differential cross sections measurements from the CMS experiment are presented. These measurements are important to test QCD predictions for the heaviest quark, to find limits of QCD applicability, as well as to search for physics beyond Standard Model as it could demonstrate itself through subtle variations in kinematic distributions. For now, as it is seen from Fig. 22 and other top quark cross sections measurements, all results are in good agreement with the Standard Model predictions.

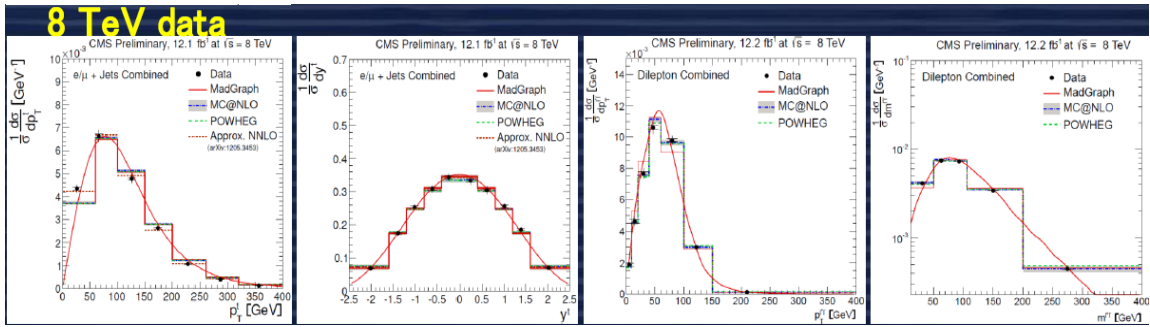

Figure 22. Top quark differential distributions at 8 TeV center of mass energy LHC.

In addition to the pair production of the top quarks via strong interaction, electroweak production of the top quarks at the LHC has sizable cross sections. Both ATLAS and CMS are now measured t-channel single top quark production with high precision (Fig. 23[29]) and extracted direct limits on the value of CKM matrix element $V_{tb}$. All results for the single top quark production are in agreement with the Standard Model predictions and stringent limits on model-independent value of $V_{tb}$ matrix element are set: $V_{tb}$ value differs from 1.0 by less than ~10%.

|  | L[fb$^{-1}$] | t-channel cross section [pb] | $|V_{tb}|$ |
|---|---|---|---|
| ATLAS | 5.8 | $95.1 \pm 2.4(\text{stat.}) \pm 18.0(\text{syst.})$ | $1.04^{+0.10}_{-0.11}$ |
| CMS | 5.0 | $80.1 \pm 5.7(\text{stat.}) \pm 11.0(\text{syst.}) \pm 4.0(\text{lumi.})$ | $0.96 \pm 0.08(\text{exp.}) \pm 0.02(\text{th.})$ |

Figure 23. Single top quark production cross sections in t-channel at 8 TeV LHC. Standard Model prediction is 89 pb with ~3% uncertainty.

An interesting puzzling effect in the top quark pair production at the Tevatron continues to attract attention of both experimentalists and theorists. Anomalous forward-backward asymmetry was observed by both CDF and D0 experiments: more top quarks follow the direction of the proton, then expected in the Standard Model. Both Tevatron experiments observe ~2σ effect and both see deviation in the same direction. No similar effect has been seen at the LHC by ATLAS and CMS, while top pair production mechanisms at the LHC are quite different. CDF experiment presented updated results on the full Tevatron data set on $A^l_{FB}$ (asymmetry calculated using leptons from decays of W bosons coming from the decays of top quarks) in lepton+jets decay channel as presented on Fig. 24[30].

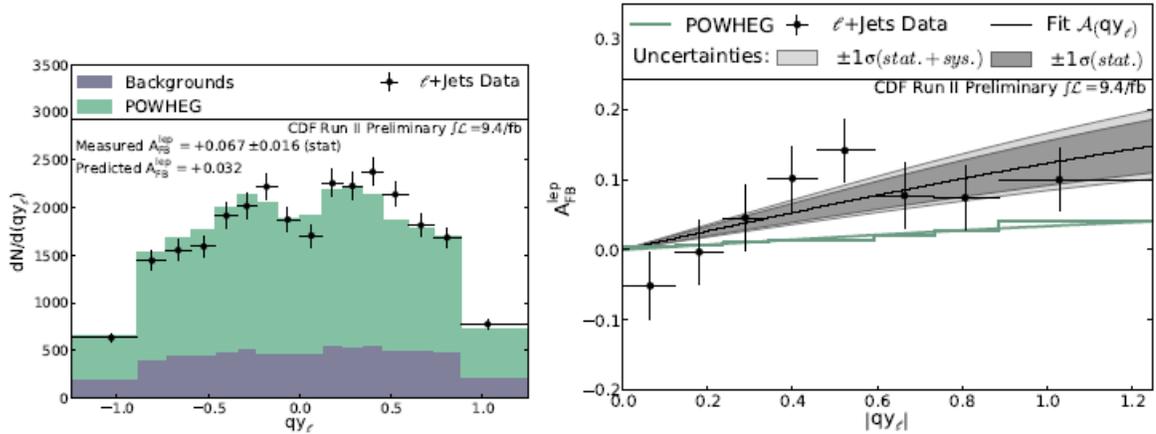

Figure 24. $A^l_{FB}$ measurements by the CDF experiment in lepton+jets channel.

After corrections for acceptance $A^l_{FB}$ is 9.4+-2.4(stat)+2.2/-1.7(syst)% compared to the theoretical prediction of 3.6% (including electroweak corrections) or ~2σ away. In order to address this puzzle results on the full Tevatron data set from the D0 experiment as well as higher accuracy results from the LHC are needed.

**7 Higgs Boson**

Discovery of a Higgs boson, particle responsible for electroweak symmetry breaking, last summer is among most important discoveries in the particle physics. Now with new particle observed main efforts are concentrated on studies of its properties, such as measurement of its mass, as well as verifying that this particle has indeed parameters predicted by the Standard Model, including spin and parity. Large number of new and updated results from both Tevatron and LHC experiments have been presented at the conference.

Tevatron experiments presented update of the Higgs boson searches on the full Tevatron 2 TeV data set of 10 fb$^{-1}$. Both Tevatron experiments, CDF and D0, see excess of events at ~2σ level. When all search channels from both experiments, including decays to bb, WW, γγ and ττ, significance of the excess at a mass of 125 GeV is 3.1σ[31]. Fig. 25a presents combined background p-value (or probability of background to fluctuate to or above number of events observed in data) demonstrating very low probability that signal observed at the Tevatron could be explained by the background fluctuation. As presented on Fig. 25b excess of events above background is observed in all studied Higgs decay channels, including decay into a pair of fermions. All Tevatron results are in agreement with the Standard Model Higgs boson production and decay adding important information to our understanding of the Higgs boson produced in conditions different than at the LHC.

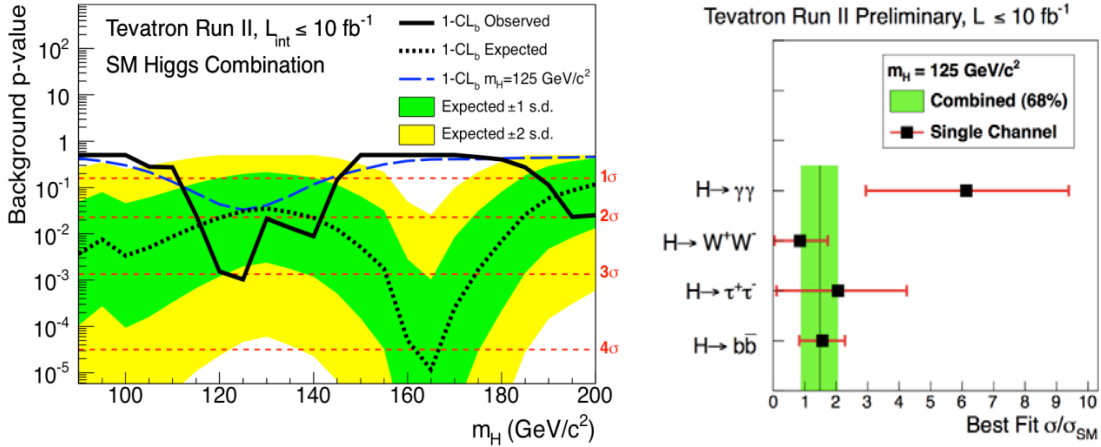

Figure 25. Background p-value for combined Tevatron Higgs boson searches result (left, 25a). Best fit Higgs boson production cross sections normalized to the Standard Model predictions for the Higgs decay channels studied at the Tevatron (right, 25b).

LHC experiments presented wealth of new experimental results about Higgs boson at the winter conferences 2013 in most cases using full 2013 8 TeV data set where studies now, with very high significance of Higgs observation already achieved, concentrate on precision measurements of its properties. We will start from discussing studies of the Higgs boson decay modes. For the decay to a pair of b-quarks[32], most probable decay mode of the Higgs boson with a mass of ~125 GeV, CMS reported 2.2σ observed excess with 2.1σ expected, providing a strong indication of the Higgs boson decaying to a pair of quarks. ATLAS expected sensitivity in this decay channel is 1.9 times above the Standard Model with observed limit 1.8 times above predictions not yet providing significant sensitivity for studies in this channel. With full 2013 data set, improvements in the analysis methods and potentially combining LHC and Tevatron experiments increases in sensitivity are expected with currently available data sets.

Decays of the Higgs boson to the heaviest lepton, τ-lepton, are searched by both ATLAS and CMS experiments. Using close to the full available data set observed(expected) significances of excess are[32]: ATLAS 1.1σ (1.7σ), CMS 2.9σ (2.9σ). CMS reached Standard Model sensitivity and for all practical purposes established evidence for the Higgs boson decay to a pair of τ-leptons. Both ATLAS and CMS results are compatible with Standard Model Higgs boson decays to τ-leptons.

One of the most stringent tests if a Higgs boson discovered is indeed particle as predicted by the Standard Model is the ratio of production cross section times branching fraction to the Standard Model predicted values. Such ratio is called μ. Both ATLAS and CMS measured μ values for a large number of different Higgs production and decay processes as presented on Fig. 26[33,34].

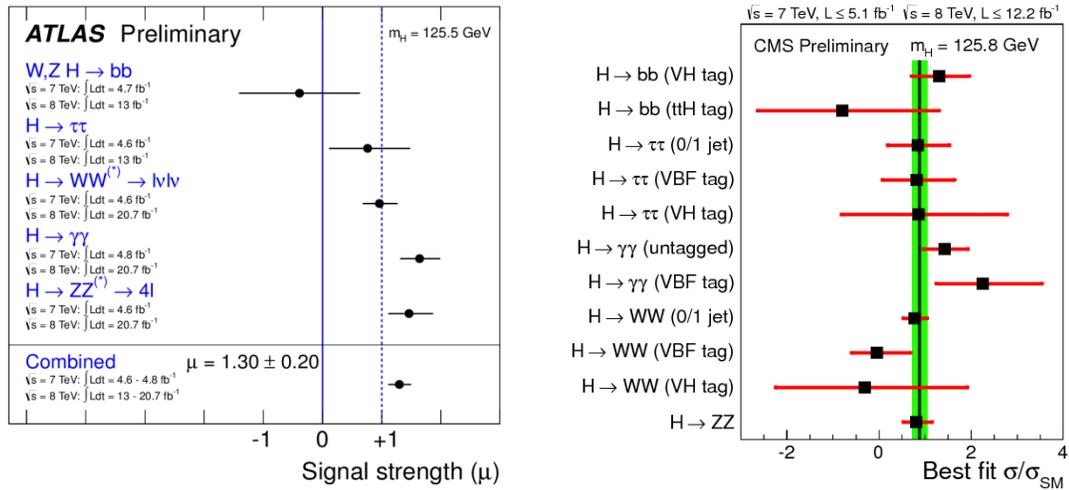

Figure 26. μ values for different Higgs boson production and decay modes by ATLAS and CMS experiments. Combined μ value for ATLAS 1.30+-0.20 and for CMS is 0.88+-0.21 (before update presented on Fig. 27).

Both ATLAS and CMS results are compatible with Standard Model, while some tension for Higgs to γγ decay existed before the conference as both experiments saw somewhat enhanced values of μ. My un-official combination of ATLAS and CMS measurements provides μ value of 1.1+-0.15 in good agreement with the Standard Model. CMS experiment updated their boson decay modes results during the conference[35] and updated CMS plot for bosonic decay channels is presented on Fig. 27. Excess originally observed in this channel has disappeared with more data and improved analyses and results are in a good agreement with the Standard Model predictions.

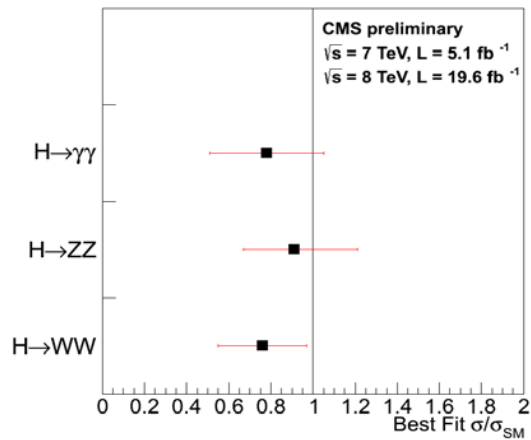

Figure 27. Updated μ values from CMS experiment on full 2013 data set.

The only parameter of the Higgs boson in the Standard Model (all other properties are predicted) is its mass. As a result precision measurements of the Higgs boson mass are very important, if not the most important measurements after a Higgs boson discovery. Both ATLAS and CMS presented full 2013 data set results on the mass of the Higgs boson. Decay channels used for the mass measurements are decays to γγ and ZZ where energy and angles of the decay products (photons and leptons) could be measured with high precision. In fact designs of both ATLAS and CMS detectors were optimized to detect Higgs boson and measure its mass with high precision and these efforts are now paying off well. Fig. 28[35] presents CMS results of measurements of the Higgs boson mass. Both channels provide consistent results with average mass of 125.8+-0.4(stat)+-0.4(syst) GeV[34].

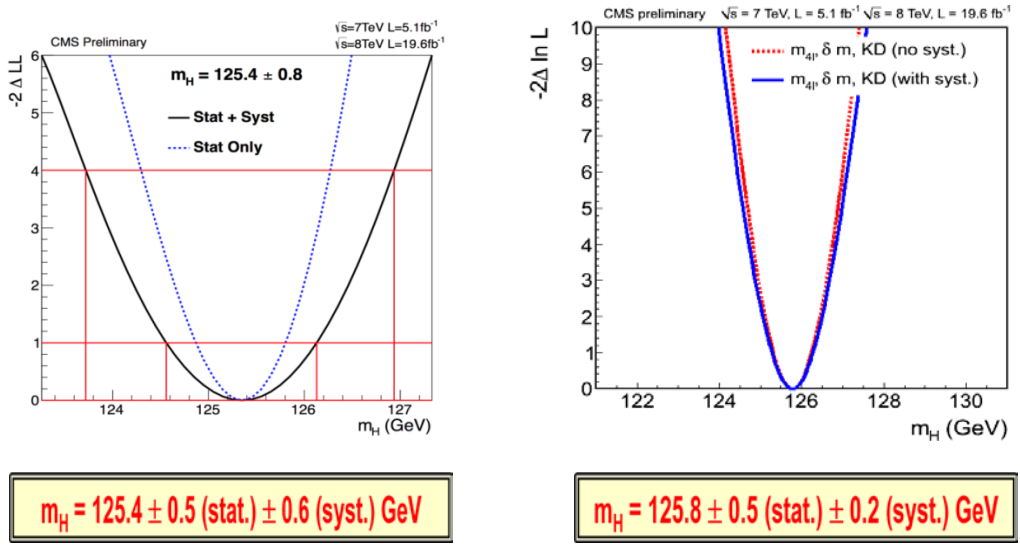

Figure 28. CMS experiment Higgs boson mass measurements in γγ (left) and ZZ (right) decay channels.

Fig. 29[33] presents ATLAS results of the Higgs boson mass measurement. Small tension between these two results exists: γγ channel best fit mass is somewhat above mass measured in ZZ channel. These two results are compatible at 2-8% level which provides an opportunity for the ATLAS experiment to combine these two results and obtain Higgs boson mass value of 125.5+-0.2(stat)+0.5-0.6(syst) GeV in a good agreement with CMS mass measurement. It is too early to say where the difference between Higgs masses measured in two different channels by the ATLAS experiment is coming from. Using existing data set with improvements in the detector calibration might reduce the tension (or enhance it?). It could be that with more data (to be collected in 2015 and beyond) this difference will go down, while there is a chance there is yet un-known physics which affects measured values in these two channels. In any scenario it is critical to continue to improve Higgs boson mass measurement by both ATLAS and CMS experiments.

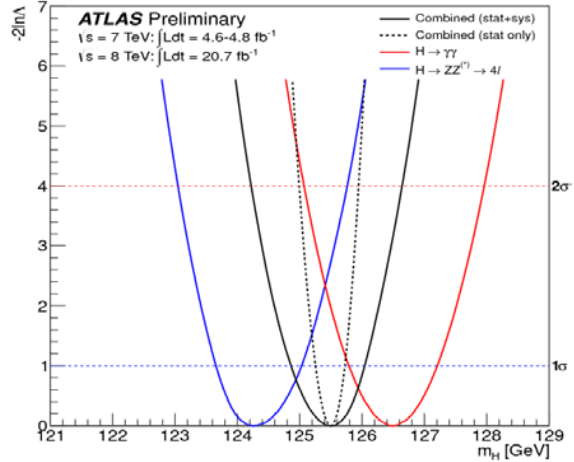

Figure 29. ATLAS experiment Higgs boson mass measurements in γγ and ZZ channels.

Back of an envelope combination of ATLAS and CMS Higgs boson mass results provides mass of 125.6+-0.3GeV. The accuracy of this result, 0.24%, is already impressive – less than a year after the Higgs boson discovery. For both experiments mass accuracy is limited by systematic uncertainties, so improvements in mass measurement methods are needed to improve precision.

Measuring spin and parity of the newly discovered particle is critical to assure that signal observed is indeed coming from the Standard Model Higgs boson ($0^+$). Both ATLAS and CMS presented extensive results on this topic. Separation between different states is mainly coming from low background, high angular resolution ZZ Higgs decay mode with Z's decaying to leptons. Fig. 30[34,36] presents some of the observed results. There is already very good separation between Standard Model prediction and exotic states, such as $0^-$, excluding this state with over 3σ significance. There are strong exclusion limits set by ATLAS and CMS for other spin-parity states.

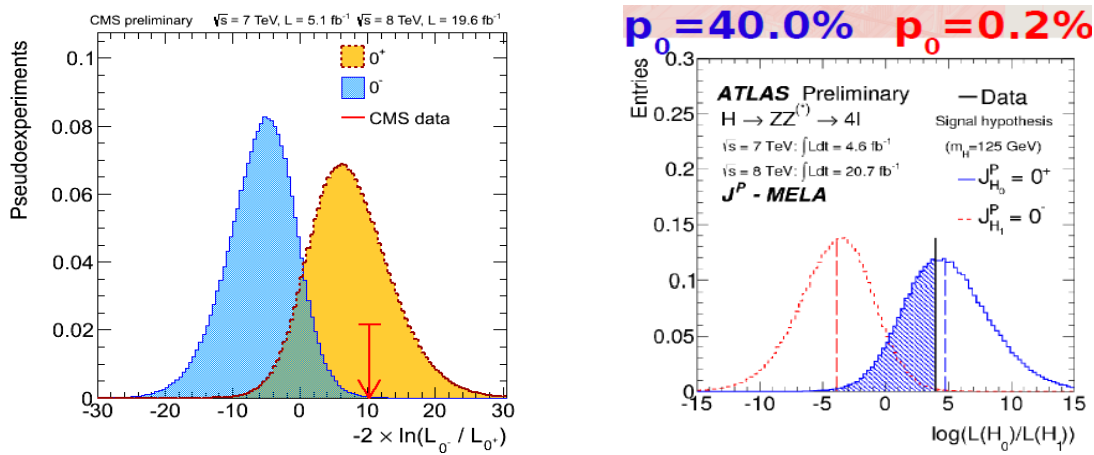

Figure 30. CMS (left) and ATLAS (right) exclusion plots for a particle with spin parity $0^-$ vs Standard Model Higgs boson prediction of $0^+$.

Another fundamental property of the Higgs boson is couplings to other particles which are expected to be proportional to the mass of a particle. The approach here is to measure couplings, normalized to their Standard Model values, $K_i$, using both production and decay rates. As different couplings might be involved in the production and decays of the Higgs boson, there are correlations between limits on different couplings. Fig. 30[33,34] presents results on the couplings obtained by the ATLAS and CMS experiments. Sub-script "F" means fermion couplings and "V" means vector boson couplings.

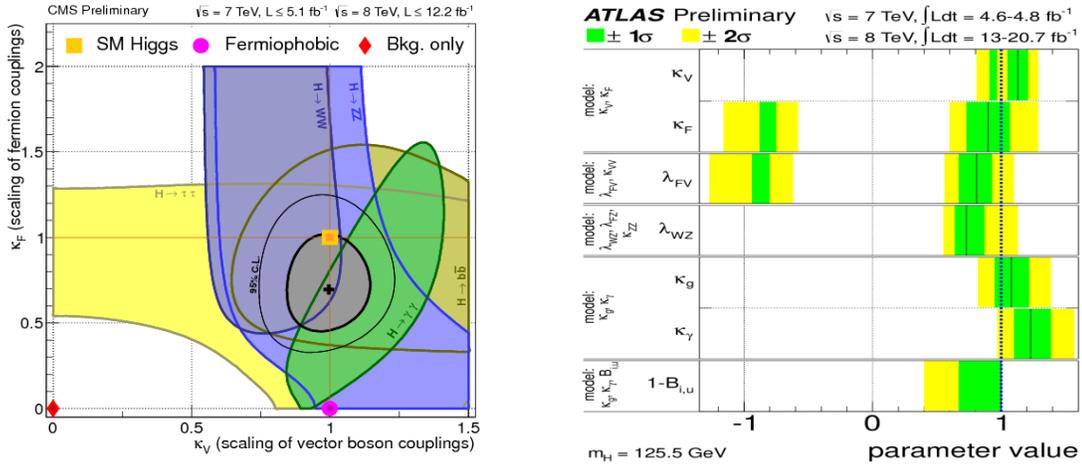

Figure 30. Higgs boson couplings measurements by CMS(left) and ATLAS(right) experiments.

All couplings results obtained so far are in good agreement with the expected in the Standard Model, while uncertainties are on a larger size: ~30% for $K_F$ and ~20% for $K_V$. Measurements of couplings will continue to improve as more data will be accumulated at the LHC and analysis methods will improve.

With all Higgs boson studied properties matching well Standard Model predictions the particle discovered in July 2012 looks more and more as the Standard Model Higgs boson. Ultimate test is the verification of the self-consistency of the Standard Model using relation between masses of the Higgs boson, the top quark and the W boson and other electroweak parameters. Gfitter collaboration[37] presented at the conference such self-consistency check with top quark and W boson masses from the Tevatron and Higgs boson mass as measured at the LHC. The plot of such a comparison is presented on Fig. 31.

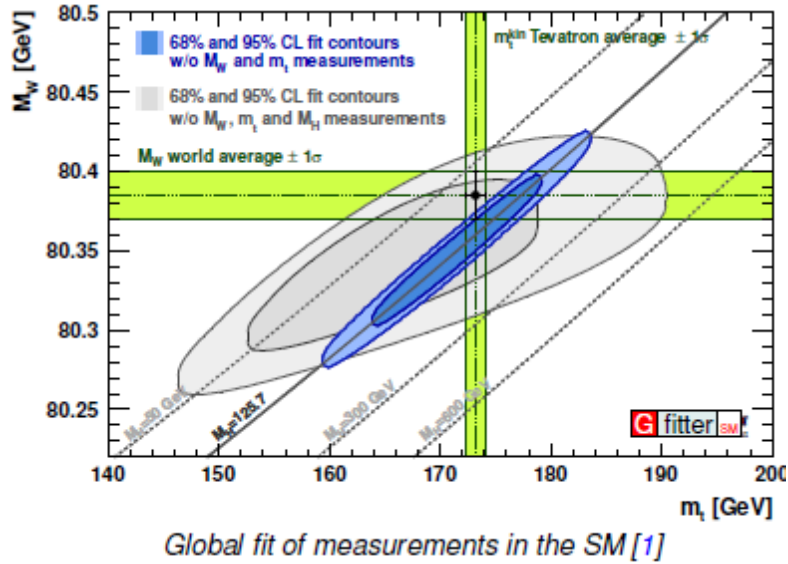

Figure 31. Standard Model self-consistency check in the top quark mass vs W boson mass plane.

The goodness of the fit is ~8% demonstrating impressive self-consistency of the Standard Model and supporting the hypothesis that the newly observed particle is indeed the Higgs boson of the Standard Model. Improvements in the measurements of the top quark, W boson and the Higgs boson masses are critical for continuing improvements of our understanding of the Standard Model applicability.

**8 Future LHC Running**

Many of the key topics of modern particle physics depend heavily on data produced by the LHC. In the talk[38] status and plans of the LHC accelerator complex for the next run, to be started in early 2015, are summarized as presented on Fig. 32. With excellent LHC performance during Run I in 2010-2013 there is high confidence that next running period plans will be realized. The center of mass energy of the collider will be in the range 13-14 TeV with ~150 days per year of delivering pp luminosity to the experiments. With planned 25ns bunch spacing in 2015 integrated luminosity per year will be in 25-45 fb$^{-1}$ range (for ATLAS and CMS experiments) with number of pile-up interactions of ~25-50. There are different options how to optimize delivered luminosity while keeping number of pile-up interactions reasonably low. The final strategy is expected to be developed in close cooperation between machine experts and LHC experiments.

| | Number of bunches | Bunch intensity [$10^{11}$ ppb] | β*/ crossing angle | Emittance LHC [μm] | Peak Luminosity [cm$^{-2}$s$^{-1}$] | ~Pile-up | Int. Lumi per year [fb$^{-1}$] |
|---|---|---|---|---|---|---|---|
| 25 ns | 2760 | 1.15 | 55/189 | 3.75 | $0.93 \times 10^{34}$ | 25 | ~24 |
| 25 ns BCMS | 2520 | 1.15 | 45/149 | 1.9 | $1.7 \times 10^{34}$ | 52 | ~45 |
| 50 ns | 1380 | 1.6 | 42/136 | 2.5 | $1.6 \times 10^{34}$ level to $0.8 \times 10^{34}$ | 87 level to 44 | ~40* |
| 50 ns BCMS | 1260 | 1.6 | 38/115 | 1.6 | $2.3 \times 10^{34}$ level to $0.8 \times 10^{34}$ | 138 level to 44 | ~40* |

Figure 32. LHC parameters for the next run planned to start in early 2015.

**10 Concluding Remarks**

With 56 excellent experimental talks presented at the conference it is impossible to do justice to all of them in the relatively brief summary. I recommend to all readers to study all of the submitted contributions presented in these proceedings. Main experimental highlights of the conference (see talk[39] of Michelangelo Mangano for the conference Theoretical Summary) based on the new results presented at Moriond QCD 2013 include

a) New results from the LHCb experiment do not confirm anomalously large CP violation in the charm sector;
b) Unexplained double-ridge structure seen by ALICE experiment in charged particles correlations in pPb collisions;
c) No *new* hints of beyond Standard Model effects in direct or indirect searches;
d) All measured Higgs boson properties match Standard Model Higgs boson very well;
e) Mass of the Higgs boson is 125.6+-0.3 GeV.

*Standard Model stands stronger than ever* indicating its fundamental character and importance of even deeper studies of one of the most successful theories of Nature.

I would like to thank all speakers of the conference for presenting excellent talks and the experimental collaborations for the large number of new, exciting, and sometime puzzling results. It was great pleasure to work closely with Michelangelo Mangano on the conference summary talks. I thank the Moriond QCD 2013 organizers for the excellent organization of the conference and for the invitation to give experimental summary talk which was challenging while exciting experience.